\documentclass[preprint,tightenlines,superscriptaddress,prd,nofootinbib,eqsecnum]{revtex4-1}
\usepackage[pdftex]{graphicx}
\usepackage{subfigure}
\usepackage{amstext}
\usepackage{amsmath}
\usepackage{endnotes}
\let\footnote=\endnote
\usepackage{amssymb}
\def\ee{{\rm e}}

\def\_#1{^{}_{#1}}

\def\be{\begin{equation}}
\def\ee{\end{equation}}
\def\bea{\begin{eqnarray}}
\def\eea{\end{eqnarray}}
\usepackage{enumerate}

\begin{document}

\title{The 2D surfaces that generate Newtonian and general relativistic orbits with small eccentricities}

\author{Chad A. Middleton}
\email{chmiddle@coloradomesa.edu}
\affiliation{Department of Physical and Environmental Sciences, Colorado Mesa University, Grand Junction, CO 81501, U.S.A.}

\begin{abstract}
Embedding diagrams prove to be quite useful when learning general relativity as they offer a way of visualizing spacetime curvature through warped two dimensional (2D) surfaces.  In this manuscript we present a different 2D construct that also serves as a useful conceptual tool for gaining insight into gravitation, in particular, orbital dynamics - namely the cylindrically symmetric surfaces that generate Newtonian and general relativistic orbits with small eccentricities.  Although we first show that no such surface exists that can \textit{exactly} reproduce the arbitrary bound orbits of Newtonian gravitation or of general relativity (or, more generally, of \textit{any} spherically symmetric potential), surfaces do exist that closely approximate the resulting orbital motion for small eccentricities; exactly the regime that describes the motion of the solar system planets.  These surfaces help to illustrate the similarities, as well as the differences, between the two theories of gravitation (i.e. stationary elliptical orbits in Newtonian gravitation and precessing elliptical-like orbits in general relativity) and offer, in this age of 3D printing, an opportunity for students and instructors to experimentally explore the predictions made by each.


\end{abstract}\maketitle

\section{Introduction}
One hundred years ago, Albert Einstein published his general theory of relativity (general relativity or GR), which describes gravity as the warping of space and time due to the presence of matter and energy, and forever changed our conceptual understanding of one of nature's fundamental forces. Prior to GR, Isaac Newton's theory of gravitation, which describes gravity as the force of one massive object on another, reigned supreme and unchallenged for nearly a quarter of a millennium. With the advent of GR, Newtonian gravitation was reduced to that of an approximation to a more fundamental underlying theory of nature and must be abandoned altogether when probing phenomena near highly relativistic objects such as black holes and neutron stars. Nevertheless, Newtonian gravitation is highly successful in accounting for the approximate motion of the planets in our solar system and allows for a theoretical derivation of Kepler's three empirical laws.

According to Newtonian gravitation (and Kepler's first law of planetary motion) the planets move in \textit{closed} elliptical orbits with the Sun residing at one focus.\cite{bertrand} For these orbits, the translating body finds itself at precisely the same radial distance after the polar angle has advanced by $2\pi$ and will therefore retrace its preceding elliptical motion indefinitely.  This Newtonian prediction of closed elliptical orbits holds true only in the absence of other massive objects as additional orbiting bodies generate perturbations to the motion. Additionally, when the theoretical framework of GR is employed, the closed or \textit{stationary} elliptical orbits of Newtonian gravitation are replaced with elliptical-like orbits whose perihelia and aphelia \textit{precess}, even in the absence of additional massive objects.\cite{perihelion}  Before Einstein's GR, the small but observable precession of Mercury's perihelion was partially unaccounted for and offered the first real test of general relativity. In fact, when the precession of Earth's equinoxes and the gravitational effects of the other planets tugging on Mercury are accounted for, GR claims exactly the additional measured precession. The predicted value of Mercury's perihelion precession due to general relativistic effects is incredibly tiny, equating to an angular shift per orbit of less than one part in ten million.

According to general relativity, a massive object causes the spacetime in its vicinity to warp or curve simply due to its presence. Hence, the planets move in their respective orbits not due to the gravitational force of the Sun acting on them, but rather due to the fact that the planets are moving in the warped four-dimensional spacetime about the Sun.  In a typical introductory undergraduate GR course, the spacetime external to a non-rotating, spherically symmetric massive object is often extensively studied as it offers a simple, exact solution to the field equations of general relativity.  To gain conceptual insight into this non-Euclidean spacetime, an embedding diagram is often constructed.  By examining the spacetime geometry external to the massive object at one moment in time, with the spherical polar angle set to $\pi/2$, a two dimensional (2D) equatorial ``slice" of the three dimensional space is extracted.  A curved two dimensional, cylindrically symmetric surface embedded in a flat three dimensional space is then constructed with the same spatial geometry as this two dimensional spacetime slice; this is the embedding diagram. These embedding diagrams prove incredibly useful in visualizing spacetime curvature and serve as a conceptual tool.

When free particle orbits, or geodesics, about non-rotating, spherically symmetric massive object are then later studied, a conceptual analogy of a marble rolling on a warped two-dimensional surface is often drawn to represent the orbital motion.  Although this analogy may be useful for the beginning student of GR in visualizing particle orbits about massive objects in curved spacetime, it isn't precise and fundamentally differs from the analogy of embedding diagrams discussed above.   Although the function that describes the shape of the cylindrically symmetric surface of the embedding diagram is of a relatively simple form, there exists no 2D cylindrically symmetric surface residing in a uniform gravitational field that can generate the exact Newtonian orbits of planetary motion, save the special case of circular orbits.\cite{english}  This null result holds when considering a fully general relativistic treatment of orbital motion about a non-rotating spherically symmetric massive object.\cite{middleton}  For an object rolling or sliding on the interior of an inverted cone with a slope of $\sqrt{2}$, it has been shown that the stationary elliptical orbits of planetary motion can occur when the orbits are nearly circular.\cite{white}  Later, elliptical-like orbits with small eccentricities were considered on 2D cylindrically symmetric surfaces where a perturbative solution was constructed.\cite{nauenberg} There it was found that when the surface's shape takes the form of a power law, precessing elliptical-like orbits with small eccentricities can only occur for certain powers and stationary elliptical orbits with small eccentricities can only occur for certain powers and for certain radii on the surface.

Although we show that no such surface exists that can \textit{exactly} reproduce the arbitrary bound orbits of a spherically-symmetric potential, cylindrically symmetric surfaces \textit{do} exist that yield the stationary and precessing elliptical-like orbits with small eccentricities of Newtonian gravitation and general relativity, respectively.  First, we arrive at a general expression describing the slope of a 2D cylindrically symmetric surface that yields the elliptical orbits of Newtonian gravitation for small eccentricities. The slope of this surface reduces to the special case of White's when the arbitrary constant of integration is set equal to zero.\cite{white} We then extend our study to general relativity and arrive at an expression for the slope of a 2D surface that generates the precessing elliptical orbits of GR for small eccentricities.  For the sake of brevity we symbolically represent \textit{stationary elliptical-like orbits of Newtonian gravitation with small eccentricities} as N$\varepsilon$ orbits and \textit{precessing elliptical-like orbits of general relativity with small eccentricities} as GR$\varepsilon$ orbits; these phrases appear frequently throughout the paper and warrant this use of shorthand.

This paper is outlined as follows. In Sec. \ref{NG}, we derive the equations of motion for an object orbiting in a generic spherically symmetric potential.  We examine the case of Newtonian gravitation and present the solution for the respective particle orbits while introducing the notation contained in the sections that follow. In Sec. \ref{Elliptical-like}, we consider the rolling or sliding motion of an object constrained to reside on a 2D cylindrically symmetric surface and obtain the corresponding orbital equation of motion. We show that this equation is fundamentally different than the equation of motion for an object subjected to a generic 3D spherically symmetric potential and that there exists no 2D cylindrically symmetric surface that can make these equations coincide for any spherically symmetric potential, except for the special case of circular orbits. For elliptical-like orbits with small eccentricities on the cylindrically symmetric surface, we solve the orbital equation of motion perturbatively. A differential equation, evaluated at the characteristic radius of the orbital motion, emerges and relates the precession parameter of the orbit to the slope of the respective surface. By demanding that the elliptical-like orbits on the surface have a constant precession parameter and by relaxing the condition that the aforementioned differential equation be evaluated at the characteristic radius of the motion, we solve the corresponding differential equation and find the slope of the surface that yields these elliptical-like orbits with small eccentricities for all radii on the surface. We then examine the solution for the slope that generates the N$\varepsilon$ orbits. In Sec. \ref{GR}, we extend our study to general relativity. By demanding that the elliptical-like orbits on the 2D surface now mimic those of GR, we present the slope of the surface that yields these GR$\varepsilon$ orbits for all radii on the surface. We then compare these two families of surfaces with the intent of offering insight into the aforementioned theories of gravitation.

\section{The classical equations of motion for an object orbiting in a spherically symmetric potential and the solution for a Newtonian potential} \label{NG}

We begin this manuscript with a review of central-force motion and, in particular, of Newtonian gravitation. This treatment is not only instructive but also serves to introduce the notation contained within the sections that follow.  

The Lagrangian describing an orbiting body of mass $m$ in a generic spherically symmetric potential is of the form
\be\label{lagrangiangen}
L=\frac{1}{2}m\left[\dot{r}^2+r^2\dot\theta^2+r^2\sin^2\theta\dot{\phi}^2\right]-U(r),
\ee
where $r,\theta,\phi$ are spherical polar coordinates and a dot indicates a time-derivative.
As the angular momentum of an orbiting body in a spherically symmetric potential is conserved, the motion of the object, without loss of generality, can be chosen to reside in the equatorial plane.  This is done by setting $\theta=\pi/2$ in Eq. \eqref{lagrangiangen} where the Lagrangian reduces to one dependent only on the two plane polar coordinates $r$ and $\phi$.  It is noted that conservation of angular momentum is a general result of central-force motion and, for the case of planetary motion in Newtonian gravitation, equates to Kepler's second law.\cite{thornton}  

The Lagrangian of Eq. \eqref{lagrangiangen} generates two equations of motion, which take the form
\bea
\ddot{r}-\frac{\ell^2}{r^3}+\frac{1}{m}\frac{dU(r)}{dr}&=&0\label{teq}\\
\dot{\phi}&=&\frac{\ell}{r^2},\label{angmom}
\eea
where $\ell$ is a constant of the motion and equates to the conserved angular momentum per unit mass.  Here we are interested in arriving at an expression for the radial distance of the orbiting body in terms of the azimuthal angle, $r(\phi)$.  Using the chain rule and Eq. (\ref{angmom}), a differential operator of the form
\be\label{oper}
\frac{d}{dt}=\frac{\ell}{r^2}\frac{d}{d\phi}
\ee
can be constructed and used to transform Eq. \eqref{teq} into the form
\be\label{phieq}
\frac{d^2r}{d\phi^2}-\frac{2}{r}\left(\frac{dr}{d\phi}\right)^2-r+\frac{r^4}{m\ell^2}\frac{dU(r)}{dr}=0.
\ee
Equation \eqref{phieq} is the equation of motion describing the radial distance of an orbiting body as a function of the azimuthal angle in a generic spherically symmetric potential.  This equation will later be compared to the equation of motion for an object rolling or sliding on a cylindrically symmetric 2D surface.  

For an object of mass $m$ residing in a Newtonian gravitational potential, the general expression of Eq. \eqref{phieq} takes the form
\be\label{phiNew}
\frac{d^2r}{d\phi^2}-\frac{2}{r}\left(\frac{dr}{d\phi}\right)^2-r+\frac{GM}{\ell^2}r^2=0,
\ee
where $G$ is Newton's constant and $M$ is the mass of the central object.  Equation \eqref{phiNew} is usually presented in a slightly simpler form in most classical dynamics texts by adopting a change of variables of the form $u\equiv 1/r$.  This substitution is unnecessary in the present treatment and will thus be avoided.  It is noted that  Eqs. \eqref{teq} and \eqref{angmom} are technically the equations of motion for the reduced mass $\mu$ of the equivalent one-body problem, but approximately describe the motion of the particle of mass $m$ in the limit that $M\gg m$.\cite{thornton}  In this limit, we can neglect the motion of the central mass $M$ treating it as fixed, which is exactly the regime this manuscript is interested in understanding.

Equation \eqref{phiNew} can be solved \textit{exactly} and yields the well-known conic sections of the form
\be\label{rext}
r(\phi)=\frac{r_0}{(1+\varepsilon\cos\phi)},
\ee
where $r_0$ is a constant, known historically as the semilatus rectum, and $\varepsilon$ is the eccentricity of the orbit. Notice that for the eccentricity, $\varepsilon=0$ corresponds to the special case of circular motion whereas $0<\varepsilon<1$ describes elliptical orbits. Also notice that Eq. \eqref{rext} is a solution to Eq. \eqref{phiNew} when $\ell^2=GMr_0$.  This relation specifies the necessary angular momentum at a given radius, characterized by $r_0$, for circular or elliptical motion to occur.

For small eccentricities, the exact solution of Eq. \eqref{rext} can be approximated as
\be\label{rapp}
r(\phi)_{ap}=r_0(1-\varepsilon\cos\phi)
\ee
to first-order in the eccentricity. Notice that when Eq. \eqref{rapp} is taken as the solution for the radius, $r_0$ corresponds to the average radius of an elliptical orbit with small eccentricity to first-order in the eccentricity.  

The eccentricities of the planets of our solar system are quite small with an average eccentricity of $\varepsilon_{ave}=0.060$, hence Eq. \eqref{rapp} yields an excellent approximation for the radii of the orbits. As a specific example, Mercury has the largest of eccentricities with a value of $\varepsilon_M = 0.2056$. Comparing the exact and approximate radii given by Eqs. \eqref{rext} and \eqref{rapp} respectively, we find a maximum percent error of $4.2\%.$\cite{percent}  See Table \ref{tab:beta} near the end of this article for the eccentricities of the solar system planets.


\section{The classical equations of motion for an object orbiting on a cylindrically symmetric surface}\label{Elliptical-like}

The Lagrangian describing a body of mass $m$ residing on a cylindrically symmetric surface in a uniform gravitational field is of the form
\be\label{lagrangian}
L=\frac{1}{2}m\left[\dot{r}^2+r^2\dot{\phi}^2+\dot{z}^2\right]+\frac{1}{2}I\omega^2-mgz,
\ee
where we have included both the translational and rotational kinetic energy of the orbiting body and the gravitational potential energy of the object at a height $z$.  
In an earlier version of this manuscript, the rolling and sliding motions of an orbiting body were collectively analyzed and presented.  In the analysis the rolling object was assumed to roll without slipping and the approximation that the angular velocity of a rolling object is proportional to the translational velocity of the object's center of mass was adopted.  There it was found that the general solution for the slope of the surface, which will be presented later in this section, does not depend on the intrinsic spin of the orbiting body.  In fact, none of the results presented in this manuscript depend on the intrinsic spin of the orbiting body. With the benefit of hindsight and for the sake of simplicity, we forego this more detailed analysis as the results are left unaltered and set the rotational kinetic energy term of Eq. \eqref{lagrangian} to zero. 

As we are interested in studying the motion of an orbiting body on a cylindrically symmetric surface, we note that there exists an equation of constraint connecting two of the three cylindrical coordinates generically of the form $z=z(r)$.  Using this fact and setting the rotational kinetic energy term to zero, Eq. \eqref{lagrangian} can be massaged into the form
\be\label{lagrangian2}
L=\frac{1}{2}m\left[(1+z'(r)^2)\dot{r}^2+r^2\dot{\phi}^2\right]-mgz(r),
\ee 
where $z'(r)\equiv dz(r)/dr$ and we used the fact that $\dot{z}=\dot{r}z'$ via the chain rule.  This Lagrangian generates two equations of motion, which have been previously derived elsewhere and are of the form
\bea
(1+z'^2)\ddot{r}+z'z''\dot{r}^2-\frac{\ell^2}{r^3}+gz'&=&0\label{teqnmotion}\\
\dot{\phi}&=&\frac{\ell}{r^2}\label{angmom2},
\eea
where, again, $\ell$ is the conserved angular momentum per unit mass.\cite{english,middleton}

Upon comparison of Eqs. \eqref{teq} and \eqref{teqnmotion}, we note that there exists no spherically symmetric potential, $U(r)$, whose presence exists in the form of a radial derivative in Eq. \eqref{teq}, that can generate the second term of Eq. \eqref{teqnmotion}.  This is easily seen as the last term of Eq. \eqref{teq} will only generate functions of $r$ and not $\dot{r}$.  Likewise, there exists no cylindrically symmetric surface, described by $z(r)$, that can generate a term capable of canceling the $\dot{r}^2$ term in Eq. \eqref{teqnmotion} as derivatives of $z(r)$ are likewise strictly functions of $r$.  The $\dot{r}^2$ term of Eq. \eqref{teqnmotion} arises from motion in the $z$-direction on the cylindrically symmetric surface where $\dot{r}\neq 0$.   Hence, Eqs. \eqref{teq} and \eqref{teqnmotion} cannot be made to match for any spherically symmetric potential, $U(r)$, or for any 2D cylindrically symmetric surface, $z(r)$, except for the trivial case of $U'(r)=z'(r)=0$.  This disparity implies that there exists \textit{no} cylindrically symmetric surface residing in a uniform gravitational field that is capable of reproducing the exact motion of a body orbiting in \textit{any} spherically symmetric potential $U=U(r)$, except for the special case of circular motion.  This null result was first pointed out for an orbiting object subjected to the Newtonian gravitational potential of a central mass and later for the case of a body orbiting about a non-rotating, spherically symmetric massive object in general relativity.\cite{english, middleton}  For the special case of circular motion, a 2D cylindrically symmetric surface can yield the orbits of Newtonian gravitation, taken here to mean that the surface yields orbits that obey Kepler's three laws, when the shape of the surface takes the form of a $-1/r$ Newtonian gravitation potential well.  This result only holds true for circular motion as elliptical-like orbits on this surface precess for all radii and hence violate Kepler's first law.\cite{nauenberg}  See Fig. \ref{fig:NewPot} for a surface of revolution plot of $z(r)$ vs $r$ of a Newtonian gravitational potential well.

There exists no cylindrically symmetric surface residing in a uniform gravitational field capable of generating the \textit{exact} Newtonian orbits of planetary motion, except for the case of circular orbits.  This can be witnessed explicitly by inserting the solution for Newtonian orbits, given by Eq. \eqref{rext}, into Eq. \eqref{teqnmotion}.  Upon this substitution, Eq. \eqref{teqnmotion} takes the form
\be\label{NewSur}
(1-\frac{gr_0^3}{\ell^2}z')+\varepsilon(2-z'^2)\cos\phi+\varepsilon^2\left[(1-2z'^2)\cos^2\phi-r_0\sin^2\phi \;z'z''\right]-\varepsilon^3\cos^3\phi \;z'^2=0,\\
\ee
where we employed the chain rule and Eq. \eqref{angmom2} in calculating time derivatives and grouped the resultant expression by powers of the eccentricity.  In order for Eq. \eqref{rext} to be an exact solution to Eq. \eqref{teqnmotion} for arbitrary eccentricity, each of the four terms of Eq. \eqref{NewSur} must vanish uniquely.  Although the first two terms can in fact be set equal to zero, which consequently would yield the required angular momentum per unit mass for elliptical or circular motion to occur and a slope for the surface, notice that the $\varepsilon^2$ and $\varepsilon^3$ terms will not vanish for any  cylindrically symmetric surface.  For circular orbits, where the eccentricity is set to zero, Eq. \eqref{rext} represents an exact solution to Eq. \eqref{teqnmotion} when the angular momentum per unit mass obeys the expression $\ell^2=gr_0^3z'$ for a given characteristic radius and slope.


\subsection{The 2D surfaces that generate elliptical-like orbits with small eccentricities}\label{pert}

Using Eq. \eqref{oper}, Eq. (\ref{teqnmotion}) can be transformed into an orbital equation of motion of the form
\be\label{phieqnmotion}
(1+z'^2)\frac{d^2r}{d\phi^2}+(z'z''-\frac{2}{r}(1+z'^2))\left(\frac{dr}{d\phi}\right)^2-r+\frac{g}{\ell^2}z'r^4=0.
\ee
Equation (\ref{phieqnmotion}) is a non-linear equation of motion that can be solved perturbatively. For elliptical-like orbits with small eccentricities, we choose an approximate solution for the radius of the form
\be\label{rnu}
r(\phi)=\overline{r}_0(1-\varepsilon\cos(\nu\phi)),\\
\ee
where $\overline{r}_0$ and $\nu$ are parameters that are to be determined by Eq. (\ref{phieqnmotion}) and $\varepsilon$ is the eccentricity of the orbit, which will be treated as small and used as our expansion parameter.\cite{radii}  As the remainder of this paper deals \textit{solely} with elliptical-like orbits of small eccentricity, all subsequent orbits should be understood as having small eccentricities, even if not explicitly stated.  

The precession parameter, $\nu$, is understood as
\be\label{precession}
\nu\equiv\frac{2\pi}{\Delta\phi},
\ee
where $\Delta\phi$ corresponds to the angular separation between two apocenters of the orbital motion.\cite{perihelion, nauenberg}
Notice that when $\nu=1$, Eq. (\ref{rnu}) has the same functional form as that of Eq. (\ref{rapp}) and describes a \textit{stationary} elliptical-like orbit on the 2D surface where the angular separation between the two apocenters is $2\pi$.  When $\nu\neq 1$, Eq. (\ref{rnu}) describes a \textit{precessing} elliptical-like orbit on the 2D surface where $\Delta\phi\neq 2\pi$. Interestingly, orbits about non-rotating, spherically-symmetric massive objects in general relativity are found to have precession parameters that take on values \textit{less than} one. This corresponds to an angular separation of $\Delta\phi>2\pi$, where the apocenter of the elliptical-like orbit marches forward in the azimuthal direction in the orbital plane.

Inserting Eq. (\ref{rnu}) into Eq. (\ref{phieqnmotion}) and keeping terms up to first-order in $\varepsilon$, we find that Eq. (\ref{rnu}) represents an approximate solution to the orbital equation of motion when
\bea
\ell^2&=&g\overline{r}_0^3z'_0\label{zeroorderN}\\
z'_0(1+z'^2_0)\nu^2&=&3z'_0+\overline{r}_0z''_0\label{firstorderN},
\eea
where $z'_0$ and $z''_0$ are radial derivatives of $z$ evaluated at $r = \overline{r}_0$.  For clarity we note that in arriving at Eqs. \eqref{zeroorderN} and \eqref{firstorderN}, we expanded the radial derivatives of $z$ about $\overline{r}_0$ to first-order in the eccentricity through
\be\label{z}
z'(r)=z_0'+(r-\overline{r}_0)z_0''=z_0'-\varepsilon \overline{r}_0\cos(\nu\phi)z_0'',
\ee
where we used Eq. \eqref{rnu} in calculating the second equality; the second derivative $z''(r)$ was expanded in an identical fashion. 

Notice that Eq. (\ref{zeroorderN}) specifies the angular momentum per unit mass needed at a given radius, characterized by $\overline{r}_0$, on a given 2D surface for elliptical or circular motion to occur.   Equation (\ref{firstorderN}) determines the precession parameter $\nu$ for elliptical-like motion on a given 2D surface whose surface is defined by $z = z(r)$.

The method of this subsection thus far follows closely to that of an earlier work.\cite{nauenberg} There it was shown that for cylindrically symmetric surfaces with a shape profile of the form
\be\label{power}
z(r)\propto-\frac{1}{r^n},
\ee
precessing elliptical-like orbits are only allowed when $n < 2$ whereas stationary elliptical-like orbits can only occur for certain radii when $n < 1$. Interestingly when $n = 1$, which corresponds to Eq. (\ref{power}) taking on a shape profile of a Newtonian gravitational potential well, there exists \textit{no} radii that will yield the stationary elliptical orbits of Newtonian gravitation.  See Fig. \ref{fig:NewPot} for a surface of revolution plot of a Newtonian gravitational potential well.

In this manuscript we wish to find the 2D cylindrically symmetric surfaces that will generate the N$\varepsilon$ and GR$\varepsilon$ orbits.  This corresponds to finding the slope of the surface that will generate elliptical-like orbits with a constant precession parameter, $\nu$, for all radii on the surface.  To find these surfaces we relax the condition that the radius of the orbit and the radial derivatives of $z$ be evaluated at $r = \overline{r}_0$ and solve the differential equation, Eq. (\ref{firstorderN}), for the slope of the surface, where we treat the precession parameter $\nu$ as an unspecified constant.  Using this method, we obtain the general solution for the slope of the surface, which takes the form
\be\label{slope}
\frac{dz}{dr}=\sqrt{\frac{3-\nu^2}{\nu^2}}\;\left(1+\kappa r^{2(3-\nu^2)}\right)^{-1/2},
\ee
where $\kappa$ is an arbitrary constant of integration and $\nu$ is a constant precession parameter.  Equation \eqref{slope} can be integrated to yield the shape profile for the 2D surface, which takes the form of a hypergeometric function.  

A 2D cylindrically symmetric surface with a slope given by Eq. \eqref{slope} will generate elliptical-like orbits for a constant precession parameter at all radii on the surface.  Notice that the slope of Eq. \eqref{slope} becomes complex when $\nu >\sqrt{3}$.  This implies that there exists no cylindrically symmetric surface capable of generating elliptical-like orbits with a constant precession parameter greater than $\sqrt{3}$.  Using Eq. \eqref{precession}, this corresponds to a minimum angular separation between two apocenters of the orbital motion of $\Delta\phi_{min}=2\pi/\sqrt{3}\;\mbox{rad}\simeq 208^\circ$.


As previously mentioned at the beginning of this section, Eq. \eqref{slope} is left unaltered when the intrinsic spin of the orbiting body is included in the analysis.  Thus, for a 2D surface whose slope is given by Eq. (\ref{slope}), both rolling and sliding objects can undergo elliptical-like orbital motion with a constant precession parameter at all radii on the surface when the right initial conditions are imposed.

\subsection{The 2D surfaces that generate N$\varepsilon$ orbits}\label{NGorbits}

\begin{figure}[!t]
\centering
\includegraphics[width=8.5cm]{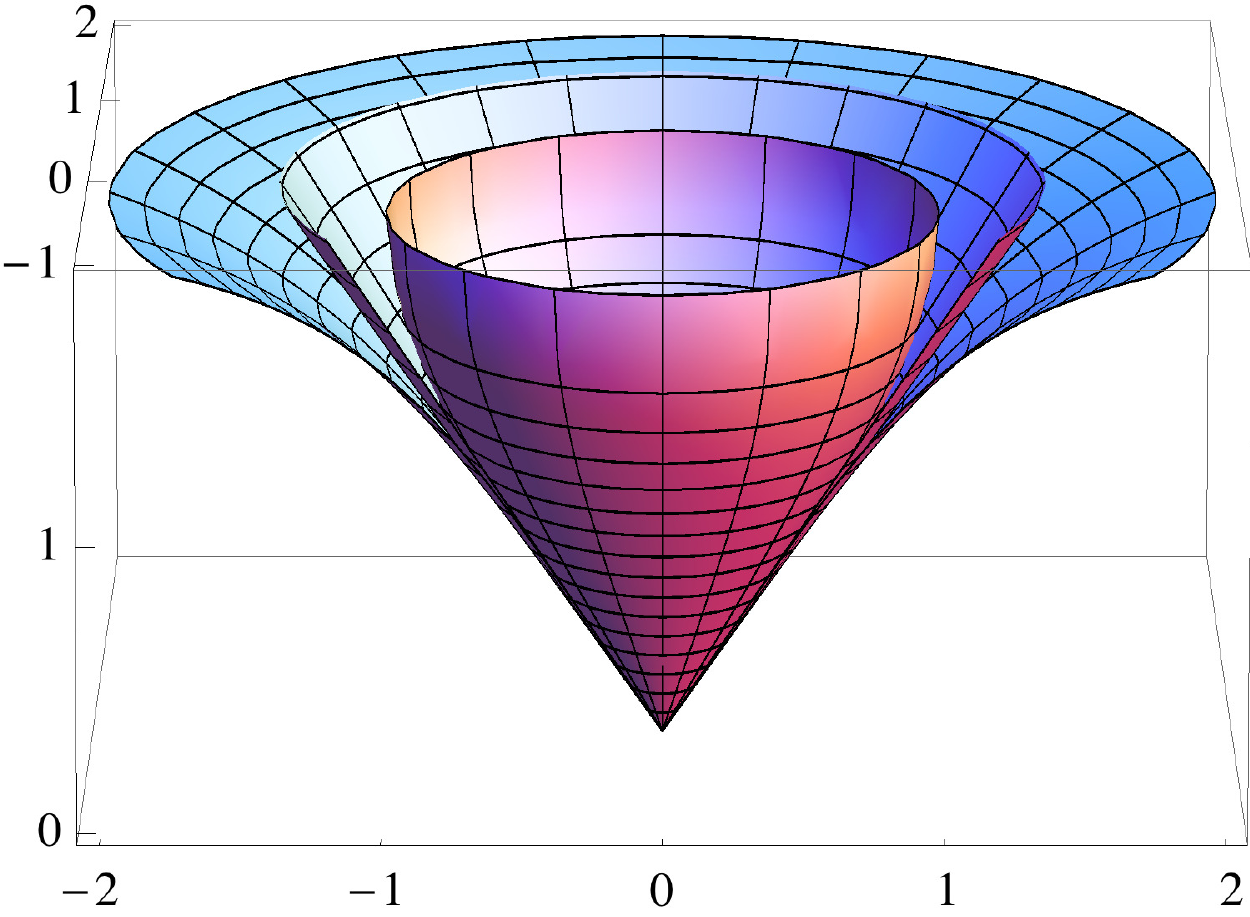}
\caption{Surface of revolution plots of $z(r)$ vs $r$ with the slope defined in Eq. (\ref{Nslope}) for $\kappa=-1,0,+1$.\cite{kappa}  Each surface will generate the stationary elliptical orbits of Newtonian gravitation when the eccentricities of the orbits are small (N$\varepsilon$ orbits).  For the $\kappa=-1$ surface, the slope of the surface diverges at $r=1$.  The $\kappa=0$ surface is that of an inverted cone with a slope of $\sqrt{2}$.  The slope of all surfaces approach $\sqrt{2}$ in the small $r$ limit.  It is noted that part of the $\kappa=0,+1$ surfaces have been removed to allow for better visibility of the $\kappa=-1$ surface.}
\label{fig:2Dplot}
\end{figure}

In this subsection, we present the 2D surfaces that yield stationary elliptical orbits with small eccentricities (N$\varepsilon$ orbits) for all radii on the surface.  Setting $\nu=1$ in Eq. \eqref{slope}, the slope of the surface takes the form
\be\label{Nslope}
\frac{dz}{dr}=\sqrt{2}(1+\kappa r^4)^{-1/2}.
\ee
Notice that for negative $\kappa$, $r$ has a range $0\leq r <(-1/\kappa)^{1/4}$, where the slope of the surface diverges at $r=(-1/\kappa)^{1/4}$.  When $\kappa$ is set equal to zero, Eq. (\ref{Nslope}) reduces to an equation of an inverted cone with a slope of $\sqrt{2}$. This result of an inverted cone with a slope of $\sqrt{2}$ giving rise to N$\varepsilon$ orbits was previously found elsewhere.\cite{white} See Fig. \ref{fig:2Dplot} for the surfaces of revolution with $\kappa > 0$, $\kappa=0$, and $\kappa<0$ that yield stationary elliptical-like orbits for all radii on its surface.  

\begin{figure}[t!]
\begin{center}
\subfigure[ Surface of revolution plot of $z(r)$ vs $r$ for a Newtonian gravitational potential well]{%
\label{fig:NewPot}
\includegraphics[width=8.5cm]{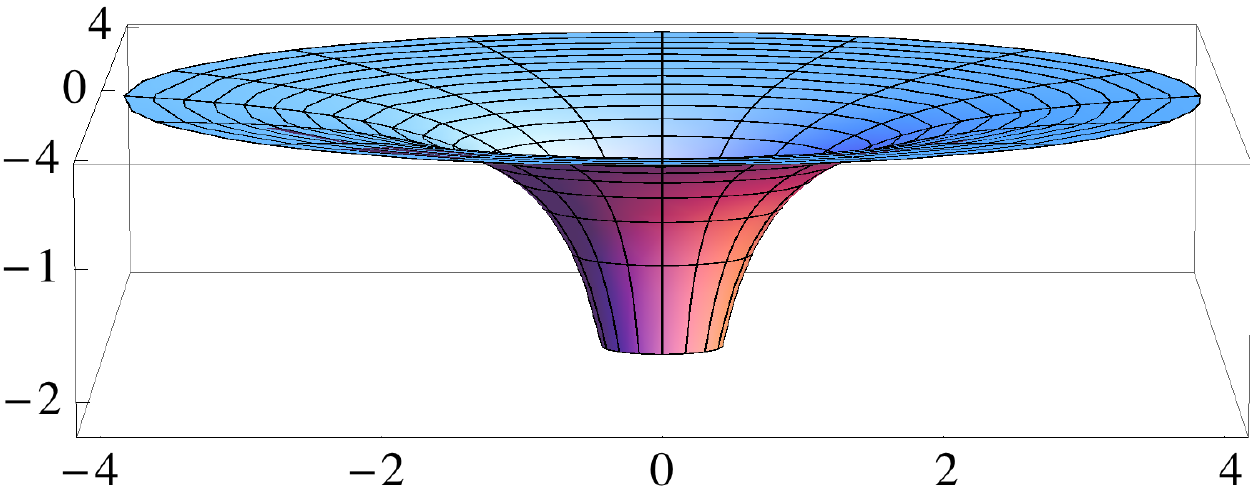}}\\%
\subfigure[Surface of revolution plot of $z(r)$ vs $r$ with the slope defined in Eq. (\ref{Nslope}) for $\kappa=+1$]{%
\label{fig:2Dplotr4}
\includegraphics[width=8.5cm]{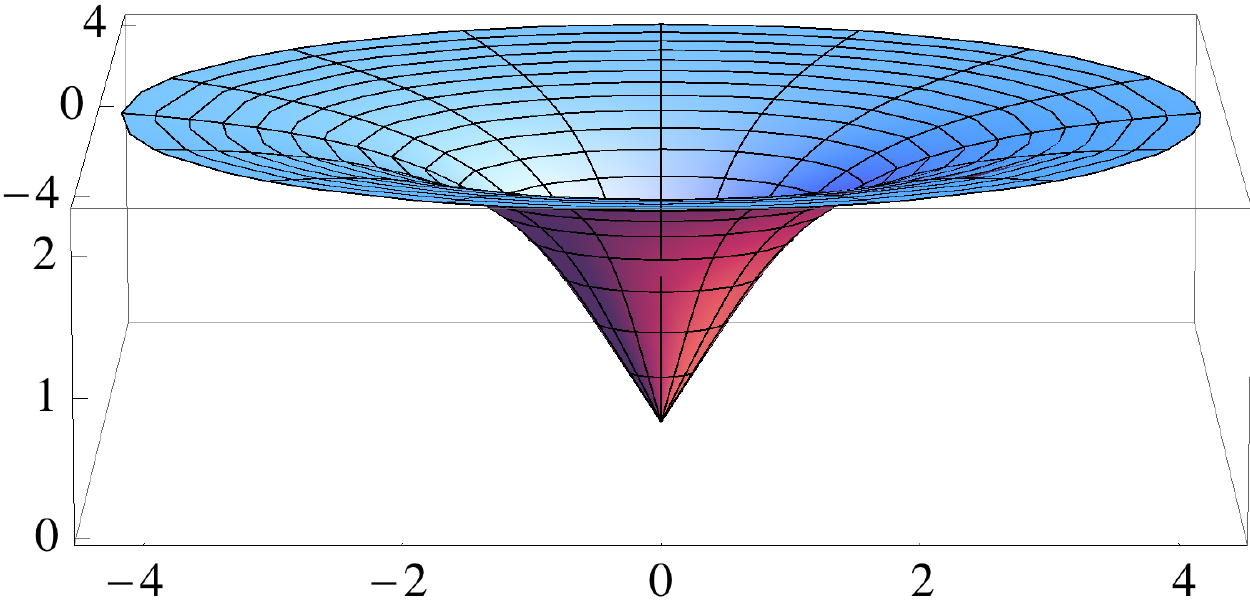}}%
\end{center}
\caption{(a) Surface of revolution plot of a $-1/r$ Newtonian gravitational potential well.  Although circular orbits on this surface obey Kepler's third law for all radial distances, there exists \textit{no} radii where the stationary elliptical orbits of Newtonian gravitation will occur.  Hence, Kepler's first law is not obeyed for orbits on this surface. (b) Surface of revolution plot with the slope defined in Eq. (\ref{Nslope}) for $\kappa=+1$.  This surface will generate the N$\varepsilon$ orbits for \textit{all} radial distances, obeying Kepler's first law to first order in the eccentricity.  Additionally, Kepler's third law is approximately obeyed for these orbits when $r\gg 1$.  Both surfaces generate particle orbits that obey Kepler's second law generically.}
\end{figure}

We further note that although Eq. (\ref{Nslope}) yields the desired N$\varepsilon$ orbits, these orbits do not, in general, obey Kepler's third law. To find the corresponding Kepler-like relation for circular orbits on a cylindrically symmetric surface with a slope given by Eq. (\ref{Nslope}), we employ Eqs. (\ref{teqnmotion}) and (\ref{angmom2}), set $\dot{r}=\ddot{r}=0$ and use the fact that $\dot{\phi}=2\pi/T$ for circular orbits. We find a relationship between the period and the radius of the orbit of the form
\be\label{period} 
T^2=\frac{2\sqrt{2}\pi^2}{\tilde{g}}r\sqrt{1+\kappa r^4}
\ee
to zeroth-order in the eccentricity. Notice that for shape profiles for the 2D surface where $\kappa r^4\ll 1$, Eq. (\ref{period}) takes on the approximate form
\be\label{periodsmall}
T^2\propto r.
\ee
This corresponds to the Kepler-like relationship for an object rolling or sliding on the inside surface of an inverted cone. Conversely, for shape profiles for the 2D surface where $\kappa r^4\gg 1$, Eq. (\ref{period}) takes on the approximate form
\be\label{periodbig}
T^2\propto r^3,
\ee
which is of course Kepler's third law for planetary motion. Notice that this $\kappa r^4\gg 1$ regime can only occur when $\kappa> 0$ as the radius of the orbiting body resides within $r <(-1/\kappa)^{1/4}$ for negative $\kappa$. Hence, for an object rolling or sliding in an elliptical-like orbit on a 2D cylindrically symmetric surface with a slope given by Eq. (\ref{Nslope}), Kepler's first law is obeyed to first-order in the eccentricity for all values of $\kappa$ and Kepler's third law is obeyed only in the $r \gg 1/\kappa^{1/4}$ regime to zeroth-order in the eccentricity. Kepler's second law is obeyed generically for central-force motion to all orders in the eccentricity.  See Fig. \ref{fig:2Dplotr4} for a surface of revolution plot of this surface.

\section{Precessing elliptical-like orbits of general relativity with small eccentricities (GR$\varepsilon$ orbits)}\label{GR}

Having found the slope of the cylindrically symmetric surface that will generate Newtonian orbits in the small eccentricity regime, we now wish to extend our study to general relativity (GR). Using Schwarzschild coordinates, the equations of motion for an object of mass \textit{m} orbiting about a non-rotating, spherically symmetric object of mass $M$ in GR are of the form
\bea
\ddot{r}-\frac{\ell^2}{r^3}+\frac{GM}{r^2}+\frac{3GM\ell^2}{c^2r^4}&=&0\label{tGR}\\
\dot{\phi}&=&\frac{\ell}{r^2},\label{angmom3}
\eea
where, in this section, a dot indicates a derivative with respect to the proper time and $\ell$ is the conserved angular momentum per unit mass.\cite{thornton,hartle,carrol}  It is noted that the above equations of motion are derivable from a relativistic Lagrangian, where here we omit the derivation.

Without loss of generality, we again choose the motion of the orbiting object to reside in the equatorial plane as the angular momentum is conserved. Upon comparison of Eqs. (\ref{tGR}) and (\ref{angmom3}) with the non-relativistic Newtonian equations of motion given by Eqs. \eqref{teq} and (\ref{angmom}) for a Newtonian gravitational potential, we note the presence of an additional term in the first equation of motion of Eq. \eqref{tGR}. This relativistic correction term offers a small-$r$ modification and gives rise to a precession of elliptical orbits, which will become apparent in what follows.

Following a similar procedure to that outlined in Section \ref{NG}, Eq. (\ref{tGR}) can be transformed into an orbital equation of motion of the form
\be\label{phiGR}
\frac{d^2r}{d\phi^2}-\frac{2}{r}\left(\frac{dr}{d\phi}\right)^2-r+\frac{GM}{\ell^2}r^2+\frac{3GM}{c^2}=0,
\ee
where we used the differential operator relation of Eq. (\ref{oper}) with the time derivative replaced with a derivative with respect to the proper time. We again wish to explore elliptical-like orbits where the eccentricities are small. We choose a perturbative solution of the form 
\be\label{rGR}
r(\phi)=r_0(1-\varepsilon\cos(\nu\phi)),
\ee
where $r_0$ and $\nu$ are free parameters that are to be determined by Eq. (\ref{phiGR}). Inserting Eq. (\ref{rGR}) into Eq. (\ref{phiGR}) and keeping terms up to first-order in $\varepsilon$, we find that Eq. (\ref{rGR}) is indeed a valid solution when the constants and parameters obey the relations
\bea
\ell^2&=&GMr_0\left(1-\frac{3GM}{c^2r_0}\right)^{-1}\label{zeroorderGR}\\
\nu^2&=&1-\frac{6GM}{c^2r_0}\label{firstorderGR}.
\eea
An expansive treatment of perihelion precession for nearly circular orbits of an object residing in a central potential and about a static spherically symmetric massive object in GR can be found elsewhere.\cite{schmidt}
Notice that when the GR correction term $GM/c^2r_0$ is set to zero, Eqs. (\ref{zeroorderGR}) and (\ref{firstorderGR}) reduce to the results of
Newtonian gravitation where $\ell^2=GMr_0$ and stationary elliptical orbits are recovered as $\nu=1$. Also notice that for a non-zero GR correction term, $\nu$ only takes on values less than one. This well known result illustrates how the stationary elliptical orbits of Newtonian gravitation are replaced with precessing elliptical orbits in GR, for orbits about a non-rotating, spherically symmetric massive object.

For a given central mass $M$, Eq. (\ref{firstorderGR}) implies that the deviation from stationary elliptical orbits increases with decreasing radius $r_0$, as should be expected. Hence, the innermost planets of our solar system experience a greater relativistic correction to the Newtonian orbits than the outermost planets. For the student of GR, one may notice that the relation presented in Eq. (\ref{firstorderGR}) is void of a dependency on the eccentricity of the orbit. In a non-perturbative treatment, the precession of elliptical orbits is in fact dependent on the eccentricity.\cite{hartle, carrol} This dependency amounts to a second-order contribution for orbits with small eccentricities and is therefore absent in our treatment as our calculation is only valid to first-order in the eccentricity.

Notice that when $r_0 < 6GM/c^2$, Eq. (\ref{firstorderGR}) yields a complex value for $\nu$. This implies that elliptical-like orbits are only allowed for radii larger than this threshold radius. This threshold radius represents the innermost \textit{stable} circular orbit (ISCO). Also notice that when $r_0 < 3GM/c^2$, the conserved angular momentum per unit mass, $\ell$, given by Eq. (\ref{zeroorderGR}) also becomes complex in addition to $\nu$. This implies that there are no circular orbits, stable or unstable, for the case of an object orbiting around a non-rotating, spherically symmetric central object with a radius $r_0 < 3GM/c^2$.

\subsection{The 2D surfaces that generate GR$\varepsilon$ orbits}\label{GRorbits}

In this subsection, we present the 2D surfaces that generate the GR$\varepsilon$ orbits.  Adopting Eq. (\ref{firstorderGR}) for $\nu$ and inserting this into Eq. \eqref{slope}, the slope of this surface takes the form
\be\label{GRslope}
\frac{dz}{dr}=\sqrt{\frac{2+\beta}{1-\beta}}\left(1+\kappa r^{2(2+\beta)}\right)^{-1/2},
\ee
where we defined the dimensionless relativistic quantity $\beta\equiv 6GM/c^2r_0$.  The validity of this solution for the slope can easily be verified by first inserting Eq. (\ref{rnu}) into Eq. (\ref{GRslope}) and expanding to first-order in $\varepsilon$ and then inserting both Eqs. (\ref{rnu}) and (\ref{GRslope}) into Eq. (\ref{phieqnmotion}), with $\nu$ defined by Eq. (\ref{firstorderGR}).

\begin{figure}[!t]
\centering
\includegraphics[width=8.5cm]{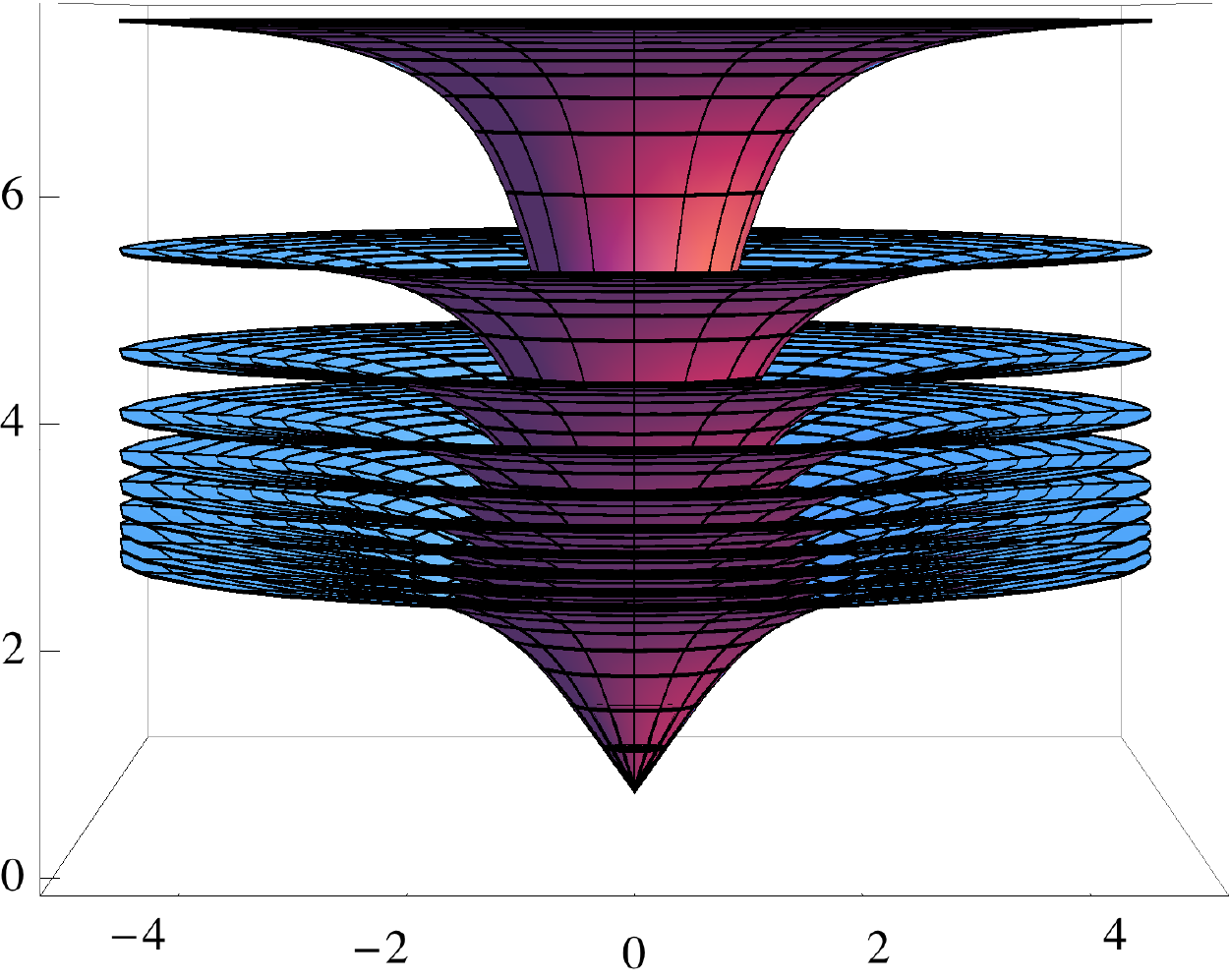}
\caption{Surface of revolution plots of $z(r)$ vs $r$ with slopes defined in Eq. (\ref{GRslope}) for $\kappa=+1$ where the relativistic quantity $\beta$ ranges from $\beta=0-0.9$ and increases in 0.1 intervals, displayed from bottom to top respectively.  Each surface will generate the \textit{precessing} elliptical-like orbits of general relativity when the eccentricities are small (GR$\varepsilon$ orbits) for the given $\beta$.  The shape of each surface, and hence the elliptical-like orbits each surface will generate, are dependent upon the dimensionless relativistic quantity $\beta\equiv 6GM/c^2r_0$.  A surface with a larger value of $\beta$, which equates to a smaller value of the precession parameter $\nu$, will generate a larger precession for the orbiting body.}
\label{fig:GRplot}
\end{figure}

A 2D cylindrically symmetric surface with a slope given by Eq. (\ref{GRslope}) will generate GR$\varepsilon$ orbits for all radii on the surface. Notice that the slope of this surface is dependent upon the central mass, $M$, and the characteristic radius of the orbiting body, $r_0$, of the celestial system whose orbital dynamics this surface seeks to mimic via the dimensionless relativistic quantity $\beta$. Further, the slope of this surface depends on $\beta$ in both the overall factor and in the power of $r$ of the 2D surface, as can be seen in Eq. (\ref{GRslope}). Notice that a surface defined by a larger $\beta$, which equates to a smaller value of the precession parameter $\nu$ via Eq. (\ref{firstorderGR}), will generate an orbit with a larger precessional rate of the apocenter, as can be seen in Eq. \eqref{precession}. Hence, one needs a unique 2D surface to replicate each unique free particle orbit in GR. This result differs dramatically from the result of the Newtonian treatment of the preceding section as the slope given by Eq. (\ref{Nslope}) is independent of the parameters of the theory.

Also notice that the general relativistic relation of Eq. \eqref{GRslope} reduces to its Newtonian counterpart given by Eq. \eqref{Nslope} when the relativistic quantity $\beta$ is set equal to zero. Although the perturbative treatment of this manuscript is only valid for orbits with small eccentricites, the treatment presented here is fully relativistic. The quantity $\beta$ has a range $0\leq\beta< 1$, as $\beta=1$ corresponds to the innermost stable circular orbit (ISCO) discussed previously where elliptical-like orbits are not allowed. As this relativistic quantity approaches one, the slope given by Eq. \eqref{GRslope} diverges for all values of $r$. See Fig. \ref{fig:GRplot} for ten stacked surface of revolution plots with $\kappa=+1$, where the relativistic quantity $\beta$ ranges over the interval $\beta= 0-0.9$, increasing in 0.1 intervals. In short, a surface defined by a larger value of $\beta$ generates an orbit with a larger precession of the apocenter.

For the planets of our solar system, the relativistic quantity $\beta$ has been calculated for each and are displayed in Table \ref{tab:beta}. We note how incredibly small the relativistic parameter is for each of the solar system planets. These values for $\beta$ offer a tiny modification to the slope of the surface defined by Eq. (\ref{Nslope}) and replace the stationary elliptical orbits of Newtonian gravitation with the precessing elliptical-like orbits of GR.

\vspace{0.5cm}
\begin{table}[t!]
\centering
\begin{tabular}{|l|c|c|c|}
\hline
Planets & \;$\varepsilon$\; & \;$r_{ave}$(m)\; & $\;\beta$\;\\
\hline
Mercury & \;0.2056\; & $5.8\times 10^{10}$ & $1.5\times 10^{-7}$\\
Venus & 0.0068 & $1.1\times 10^{11}$ & $8.3\times 10^{-8}$ \\
Earth & 0.0167 & $1.5\times 10^{11}$ & $6.0\times 10^{-8}$\\
Mars & 0.0934 & $2.3\times 10^{11}$ & $3.9\times 10^{-8}$\\
Jupiter & 0.0484& $7.8\times 10^{11}$ & $1.2\times 10^{-8}$\\
Saturn & 0.0557 & $1.4\times 10^{12}$ & $6.2\times 10^{-9}$\\
Uranus & 0.0472 & $2.9\times 10^{12}$ & $3.1\times 10^{-9}$\\
Neptune & 0.0086 & $4.5\times 10^{12}$ & $2.0\times 10^{-9}$\\
\hline
\end{tabular}
\caption{Table displaying the eccentricities of the orbits, the average Sun-planet distances, and the dimensionless relativistic quantity $\beta\equiv 6GM/c^2r_0$ for the solar system planets.\cite{data}  The small values of $\beta$ for the solar system planets reveal the scale of the modifications needed to bring the slopes of the surfaces that generate the N$\varepsilon$ orbits to coincide with those that generate the GR$\varepsilon$ orbits.}
\label{tab:beta}
\end{table}

\section{conclusion}

In this manuscript, we showed that there exists no 2D cylindrically symmetric surface residing in a uniform gravitational field that is capable of reproducing the exact particle orbits of an object subjected to a generic 3D spherically symmetric potential.  By employing a perturbative method to the equations of motion for an object orbiting on a 2D cylindrically symmetric surface, we found the general solution for the slope of the surface, given by Eq. (\ref{slope}), that will generate elliptical-like orbits with small eccentricities with a constant precession parameter for all radii on the surface.  We then examined this solution for the special cases of the surfaces that will generate N$\varepsilon$ and GR$\varepsilon$ orbits. For a 2D surface with a slope given by Eq. \eqref{Nslope} with $\kappa> 0$, N$\varepsilon$ orbits can be generated to first-order in the eccentricity, where Kepler's third law will be met to lowest order in the eccentricity when $r\gg 1/\kappa^{1/4}$.  The orbits generated on this surface differ from the elliptical-like orbits on a 2D surface corresponding to a Newtonian gravitational potential well, where Kepler's first law is not obeyed for any radius on the surface.  For a 2D surface with a slope given by Eq. \eqref{GRslope}, GR$\varepsilon$ orbits can be generated.  The slopes of these respective surfaces are functionally dependent upon the mass of the central object and the radius of the orbiting body of the gravitationally bound system whose orbital dynamics these surfaces seek to mimic. These surfaces reduce to their Newtonian counterpart when the relativistic term is set to zero, as is expected.

A comparative study of these surfaces, whose slopes are defined by Eqs. \eqref{Nslope} and \eqref{GRslope}, offers a concrete visualization of the deviations that emerge between Newtonian gravitation and general relativity for an object orbiting about a spherically symmetric massive object, when the eccentricity is small. As embedding diagrams prove quite useful to the beginning student of general relativity, it can be argued that these 2D cylindrically symmetric surfaces that yield N$\varepsilon$ and GR$\varepsilon$ orbits are equally useful in gaining insight into these respective theories of gravitation.

An interesting undergraduate physics/engineering project could entail the manufacturing and experimental testing of a 2D cylindrically symmetric surface whose slope is defined by Eq. \eqref{Nslope}.  With the advent of 3D printing, the manufacturing of such a classroom-size surface should be feasible.  Using a camera mounted above the surface and Tracker,\cite{tracker} a video-analysis and modeling software program, the angular separation between successive apocenters of elliptical-like orbits with differing eccentricities could be measured and then compared with the theoretical prediction of $2\pi$.  As the analysis presented in the manuscript is valid to first-order in the eccentricity, one expects a larger deviation from $2\pi$ for an orbit with a larger eccentricity.  For details on the above experimental setup, which was used to analyze circular orbits on a warped spandex surface, see a previous work of the author.\cite{middleton} For the student who seeks to generate the GR$\varepsilon$ orbits, a surface with a slope defined by Eq. \eqref{GRslope} could equivalently be constructed.  To generate an elliptical-like orbit that advances by $10^\circ$ per revolution, the relativistic quantity $\beta$ should be assigned a value of $\beta\simeq 0.05$ for the slope of this surface.  The inclusion of these surfaces into the classroom could offer the student of Newtonian gravitation and general relativity an incredibly useful tool, as the experimenter can see, and measure, some of the properties of these elliptical-like orbits firsthand.


\begin{thebibliography}{99}
\bibitem{perihelion}
The perihelion (aphelion) is the point of closest (farthest) approach of a planet in orbit around the Sun.  These points march forward, or precess, in the azimuthal direction when additional orbiting bodies are included in the Newtonian treatment or when the theoretical framework of GR is employed. The pericenter (apocenter) refers to the point on the elliptical orbit that is closest to (farthest from) the central mass.
\bibitem{percent}
This percent error is calculated from $|r(\phi)-r_{ap}(\phi)|/r(\phi)*100\%=\varepsilon^2\cos^2\phi*100\%$, where we used Eqs. \eqref{rext} and \eqref{rapp} in calculating the second equality.  This percent error is independent of $r_0$, depending only on the eccentricity of the orbit and the angular position of the planet, and yields a maximum value when $\phi=0,\pi$.
\bibitem{kappa}
Notice that the numerical value of $\kappa$ merely sets the scale for the 2D surface and, without loss of generality, can be normalized to take on values of $\kappa\in(-1,0,+1)$.  This result is reminiscent of the curvature constant that arises in FRW cosmology.\cite{hartle,carrol}
\end{thebibliography}
 \end{document}